\documentclass[journal,onecolumn,draftclsnofoot,11pt]{IEEEtran}

\usepackage{times}
\usepackage{epsfig}
\usepackage{graphicx}
\usepackage{amsmath}
\usepackage{amssymb}
\usepackage{graphics}
\usepackage{epsfig}
\usepackage[english]{babel}
\usepackage{doi}
\usepackage{xspace}
\usepackage{bm}
\usepackage[caption=false,font=footnotesize]{subfig}
\usepackage[utf8]{inputenc}
\usepackage[T1]{fontenc}
\usepackage{enumitem}
\usepackage{xcolor}
\usepackage{setspace} 
\usepackage{algorithm} 
\usepackage{algpseudocode} 
\usepackage{listings}
\usepackage{mathtools} 
\usepackage{tcolorbox}
\usepackage{cite}
\usepackage{environ}
\usepackage{hyperref}

\hypersetup{
    colorlinks = true,
    linkcolor = [rgb]{0.6,0,0},
    citecolor = [rgb]{0,0.6,0},
    urlcolor = [rgb]{0,0,0.6}
}

\definecolor{stringcolor}{RGB}{186,34,32}
\definecolor{digitcolor}{RGB}{0,196,0}
\definecolor{keywordcolor}{RGB}{0,128,0}
\definecolor{commentcolor}{RGB}{64,128,128}
\definecolor{operatorcolor}{RGB}{170,35,255}
\definecolor{backgroundcolor}{RGB}{245,245,246}

\definecolor{commentsColor}{rgb}{0.497495, 0.497587, 0.497464}
\definecolor{keywordsColor}{rgb}{0.000000, 0.000000, 0.635294}
\definecolor{stringColor}{rgb}{0.558215, 0.000000, 0.135316}

\definecolor{codegreen}{rgb}{0,0.6,0}
\definecolor{codegray}{rgb}{0.5,0.5,0.5}
\definecolor{codepurple}{rgb}{0.58,0,0.82}
\definecolor{backcolour}{rgb}{0.95,0.95,0.92}

\definecolor{ultramarine}{rgb}{0,32,96}

\algtext*{EndFor}

\newtcolorbox{tutorialbox}[1]{
	colback=gray!10!white,
	colframe=gray!80!black,
	arc=0mm, 
	boxrule=0.2mm, 
	fonttitle=\bfseries, 
	float=htpb!, 
	title={#1} 
}

\def\defn{\,\coloneqq\,}
\DeclareMathOperator*{\argmin}{arg min}

\newcommand{\prox}{\mathrm{prox}}

\newcommand{\R}{\mathbb{R}}

\newcommand{\E}{\mathbb{E}}

\newcommand{\Ibf}{\mathbf{I}}

\newcommand{\Xcal}{\mathcal{X}}

\newcommand{\thetabm}{\bm{\theta}}
\newcommand{\ebm}{\bm{e}}
\newcommand{\sbm}{\bm{s}}
\newcommand{\xbm}{\bm{x}}
\newcommand{\ybm}{\bm{y}}
\newcommand{\ubm}{\bm{u}}
\newcommand{\vbm}{\bm{v}}
\newcommand{\wbm}{\bm{w}}
\newcommand{\zbm}{\bm{z}}
\newcommand{\Abm}{\bm{A}}

\newcommand{\Wbm}{\bm{W}}

\begin{document}
\setlength{\arraycolsep}{0.8mm}



\title{Plug-and-Play Methods for Integrating Physical and Learned Models in Computational Imaging}

\author{
  Ulugbek~S.~Kamilov,
  Charles~A.~Bouman,
  Gregery~T.~Buzzard,
  Brendt~Wohlberg
}

\maketitle

Plug-and-Play Priors (PnP) is one of the most widely-used frameworks for solving computational imaging problems through the integration of \emph{physical models} and \emph{learned models}.
PnP leverages high-fidelity physical sensor models and powerful machine learning methods for prior modeling of data to provide state-of-the-art reconstruction algorithms. PnP algorithms alternate between minimizing a data-fidelity term to promote data consistency and imposing a learned regularizer in the form of an image denoiser. Recent highly-successful applications of PnP algorithms include bio-microscopy, computerized tomography, magnetic resonance imaging, and joint ptycho-tomography. This article presents a unified and principled review of PnP by tracing its roots, describing its major variations, summarizing main results, and discussing applications in computational imaging.  We also point the way towards further developments by discussing recent results on equilibrium equations that formulate the problem associated with PnP algorithms.

\section{Historical Background}

Consider the inverse problem of estimating an unknown image $\xbm \in \R^n$ from its noisy measurements $\ybm \in \R^m$. It is common to formulate this problem using the optimization
\begin{equation}
\label{eq:RegularizedInversion}
\widehat{\xbm} = \argmin_{\xbm \in \R^n} f(\xbm) \quad\text{with}\quad f(\xbm) = g(\xbm) + h(\xbm)\,,
\end{equation}
where $g$ is a \emph{data-fidelity term} that quantifies consistency with the observed measurements $\ybm$ and $h$ is a \emph{regularizer} that enforces prior knowledge on $\xbm$. The formulation in Eq.~\eqref{eq:RegularizedInversion} corresponds to the \emph{maximum a posteriori probability (MAP)} estimator when
\begin{equation} \label{eq:gh}
g(\xbm) = -\log(p_{\ybm|\xbm}(\xbm)) \quad\text{and}\quad h(\xbm) = -\log(p_{\xbm}(\xbm))\,,
\end{equation}
where $p_{\ybm|\xbm}$ is the likelihood relating $\xbm$ to measurements $\ybm$ and $p_{\xbm}$ is the prior distribution. For example, given measurements of the form $\ybm = \Abm\xbm+\ebm$, where $\Abm$ is the \emph{measurement operator} (also known as the \emph{forward operator}) characterizing the response of the imaging instrument and $\ebm$ is \emph{additive white Gaussian noise (AWGN)}, the data-fidelity term reduces to the least-squares function $g(\xbm) = \frac{1}{2}\|\ybm-\Abm\xbm\|_2^2$. On the other hand, many popular image regularizers are based on a sparsity promoting regularizer $h(\xbm) = \tau \|\Wbm\xbm\|_1$, where $\tau > 0$ is the regularization parameter and $\Wbm$ is a suitable transform. Over the years, a variety of reasonable choices of $h$ have been proposed, with examples including the total variation (TV) and Markov random field functions. These functions have elegant analytical forms, and have had a major impact in applications ranging from tomography for medical imaging to image denoising for cell-phone cameras.

The solution of Eq.~\eqref{eq:RegularizedInversion} balances the requirements to be both data-consistent and plausible, which can be intuitively interpreted as finding a balance between two manifolds: the \emph{sensor manifold} and \emph{prior manifold}. The sensor manifold is represented by small values of $g(\xbm)$, and in the case of a linear forward model, is roughly an affine subspace of $\R^n$. Likewise, the prior manifold is represented by small values of $h(\xbm)$ and includes the images that are likely to occur in our application. Importantly, real images have enormous amounts of structure, departures from which are immediately noticeable to a domain expert. Consequently, plausible images lie near a lower dimensional manifold in the higher dimensional embedding space.

\begin{figure}
\linespread{1}
\begin{minipage}[t]{.49\textwidth}
\begin{algorithm}[H]
\caption{ADMM}\label{alg:admm}
\begin{algorithmic}[1]
\State \textbf{input: } $\ubm^0 = \bm{0}$, $\xbm^0$, and $\gamma > 0$
\For{$k = 1, 2, \dots, t$}
\State $\zbm^k \leftarrow \prox_{\gamma g}(\xbm^{k-1} - \ubm^{k-1})$
\State $\xbm^k \leftarrow \prox_{\gamma h}(\zbm^k + \ubm^{k-1})$
\State $\ubm^k \leftarrow \ubm^{k-1} + (\zbm^k - \xbm^k)$
\EndFor
\State \Return{$x^t$}
\end{algorithmic}
\end{algorithm}%
\end{minipage}
\hspace{0.25em}
\begin{minipage}[t]{.49\textwidth}
\begin{algorithm}[H]
\caption{FISTA}\label{alg:fista}
\begin{algorithmic}[1]
\State \textbf{input: } $\xbm^0 = \sbm^0$, $\gamma > 0$, and $\{\theta_k\}_{k \geq 0}$
\For{$k = 1, 2, \dots, t$}
\State $\zbm^k \leftarrow \xbm^{k-1}-\gamma \nabla g(\xbm^{k-1})$
\State $\sbm^k \leftarrow \prox_{\gamma h}(\zbm^k)$
\State $\xbm^k \leftarrow (1-\theta_k) \sbm^k + \theta_k\sbm^{k-1} $
\EndFor
\State \Return{$x^t$}
\end{algorithmic}
\end{algorithm}%
\end{minipage}
\label{fig:minimization}
\caption{Alternating direction method of multipliers (ADMM) and fast iterative shrinkage/thresholding algorithm (FISTA) are two widely-used iterative algorithms for minimizing composite functions $f(\xbm) = g(\xbm) + h(\xbm)$ where the regularization term $h$ is nonsmooth. Both functions avoid differentiating $h$ by evaluating its proximal operator.}
\end{figure}

Proximal algorithms are often used for solving problems of the form in Eq.~\eqref{eq:RegularizedInversion} when $g$ or $h$ are nonsmooth~\cite{Parikh.Boyd2014}. One of the most widely-used and effective proximal algorithms is the \emph{alternating direction method of multipliers (ADMM)}, which uses an augmented Lagrangian formulation to allow for alternating minimization of each function in turn (see~\cite{boyd2011distributed} for an overview of ADMM).

ADMM computes the solution of Eq.~\eqref{eq:RegularizedInversion} by iterating the steps summarized in Algorithm~\ref{alg:admm} until convergence. One important property of ADMM is that  it does not explicitly require knowledge of either $g$ or $h$ or their gradients, relying instead on the \emph{proximal operator}, which is defined as
\begin{equation}
\label{eq:ProximalOperator}
\prox_{\tau h}(\zbm) \,\coloneqq \argmin_{\xbm \in \R^n} \left\{\frac{1}{2}\|\xbm-\zbm\|_2^2 + \tau h(\xbm)\right\}\,,
\end{equation}
for any any proper, closed, and convex function $h$~\cite{boyd2011distributed}. Comparing Eq.~\eqref{eq:ProximalOperator} and Eq.~\eqref{eq:RegularizedInversion}, we see that the proximal operator can be interpreted as a MAP estimator for the AWGN denoising problem
\begin{equation}
\label{Eq:MAPDenoise}
\zbm = \xbm_0 + \wbm \quad\text{where}\quad \xbm_0 \sim p_{\xbm_0}\,, \quad \wbm \sim \mathcal{N}(0, \tau \Ibf)\,,
\end{equation}
by setting $h(\xbm) = -\log(p_{\xbm_0}(\xbm))$. 

This perspective inspired the development of Plug-and-Play Priors (PnP) in \cite{venkatakrishnan2013plug}, where the $\prox_{\gamma h}$ step in ADMM is simply replaced by a more general black-box denoiser $D: \R^n \rightarrow \R^n$, such as BM3D \cite{DabovBM3D07}. That is, any black-box denoiser $D$ can in principle replace (``plug'' in for) $\prox_{\gamma h}$, and then the ADMM algorithm can run (``play'') as before.

We refer to this original algorithm as PnP-ADMM in order to distinguish it from other methods inspired by this Plug-and-Play approach.
In fact, there are multiple algorithms using proximal maps to minimize a sum of convex functions, and for each of these algorithms, there is a corresponding PnP version obtained by associating the proximal map with the prior term, then replacing the proximal map with a black-box denoiser.  
Below we provide more detail on PnP-ADMM and PnP-FISTA \cite{Kamilov-2017} (based on the proximal-gradient method), as well as extensions and variations.   See \cite{r1-beck2009fast} for the roots of FISTA, \cite{combettes2011proximal} for more detail on proximal splitting methods in general, and \cite{FCI2022} for a tutorial overview of some PnP methods.  

\begin{tutorialbox}{Turning an Image Denoising Network into an Image Super-Resolver}
PnP can be applied to multiple imaging problems using a single CNN denoiser simply by changing the physics-based measurement model.  Consider \emph{image super-resolution (ISR)} with factors $2\times$ and $4\times$, where the goal is to recover a high-resolution (HR) image from its \emph{blurred}, \emph{decimated}, and \emph{noisy} low-resolution (LR) observation. As shown below, a single DnCNN denoiser can be used within PnP to address both problems, thus leveraging the implicit image model in a deep CNN over multiple problems without retraining.

\medskip\noindent\emph{\textbf{Step 1: Learn a Denoiser.}} Let $\Xcal \subset \R^n$ denote a training dataset of natural images. The denoiser is trained by updating the weights $\thetabm$ of a deep CNN $D_{\thetabm}$ to remove the noise from $\zbm = \xbm_0 + \wbm$, where $\xbm_0 \in \mathcal{X}$ and $\wbm \sim \mathcal{N}(0, \sigma^2\bm{I})$. It is worth mentioning that since the prior in PnP is learned on a pretext task (\emph{image denoising}) rather than on the final task (\emph{image reconstruction}), PnP can be considered a \emph{self-supervised learning} framework.
\begin{figure}[H]
\centering
\includegraphics[width=13cm]{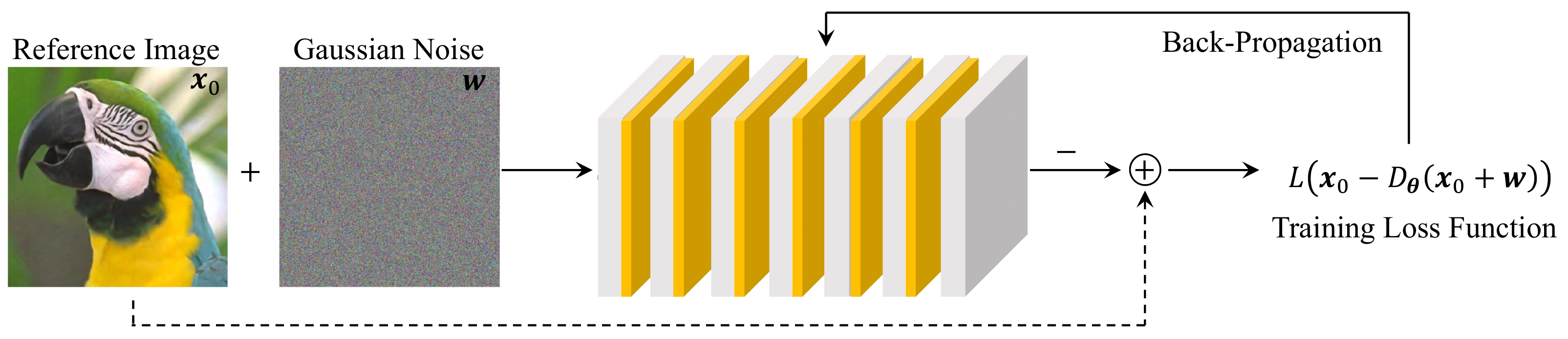}
\caption{Image priors for PnP can be obtained by training CNNs to remove AWGN from a set of images.}
\end{figure}

\vspace{-7pt}
\noindent\emph{\textbf{Step 2: Turn Denoiser into Super-Resolver.}} A pre-trained denoiser $D_{\thetabm}$ can be used for ISR by replacing $\prox_{\gamma h}$ in ADMM or FISTA by $D_{\thetabm}$. Figure~\ref{Fig:SR} shows the results obtained for two upsampling rates $2\times$ and $4\times$ using the same DnCNN denoiser, either to postprocess the pseudo-inverse or as an image prior within PnP. Note how PnP obtains significantly better results by integrating information both from the physical and learned models.

\begin{figure}[H]
\centering\includegraphics[width=12cm]{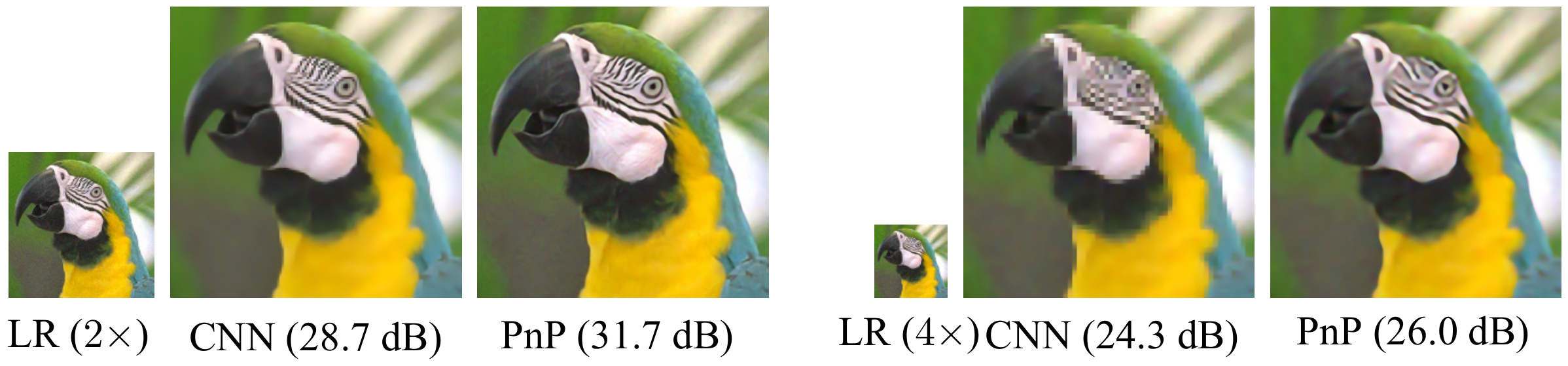}
\caption{A single pre-trained CNN denoiser in PnP can address different super-resolution factors. Code available  at \url{https://github.com/lanl/scico-data/blob/main/notebooks/superres_ppp_dncnn_admm.ipynb}}
\label{Fig:SR}
\end{figure}

\end{tutorialbox}

\section{Plug-and-Play Integration of Physical and Learned Models} \label{sec:pnp}

\emph{Deep learning (DL)} has emerged as a powerful paradigm for designing algorithms for various image restoration and reconstruction tasks, including denoising, deblurring, and super-resolution (the literature is vast, but see \cite{schmidhuber2015deep} for an early history).

Given a set of paired data $(\xbm_i, \zbm_i)$, where $\xbm_i$ is the desired ``ground truth'' image and $\zbm_i$ is its noisy or corrupted observation, the traditional supervised DL strategy is to learn a mapping from $\zbm_i$ to $\xbm_i$ by training a deep \emph{convolutional neural network (CNN)}.  Despite its empirical success in computational imaging, an important drawback of DL relative to regularized inversion is the potential need to retrain the CNN for different measurement operators and noise levels. 

The success of CNNs as black-box denoisers leads naturally to their use with PnP (see the inset \emph{``Turning an Image Denoising Network into an Image Super-Resolver''}). 
In its simplest form, PnP can be implemented by pre-training an image-denoising CNN $D$ and using $D$ in place of $\prox_{\gamma h}$ within ADMM. 
Remarkably, this simple heuristic of using denoisers not associated with any $h$ exhibited great empirical success and spurred much theoretical and algorithmic work on PnP and other related methods.  
As a result, PnP-inspired methods are among the most widely-used approaches
for combining the advantages of regularized inversion, which is flexible to changes in the data-fidelity term, with the powerful representation capabilities of deep CNNs.

\begin{figure}
\linespread{1}
\begin{minipage}[t]{.49\textwidth}
\begin{algorithm}[H]
\caption{PnP-ADMM~\cite{venkatakrishnan2013school}}\label{alg:pnpadmm}
\begin{algorithmic}[1]
\State \textbf{input: } $\ubm^0 = \bm{0}$, $\xbm^0$, and $\gamma > 0$
\For{$k = 1, 2, \dots, t$}
\State $\zbm^k \leftarrow \prox_{\gamma g}(\xbm^{k-1} - \ubm^{k-1})$
\State $\xbm^k \leftarrow D(\zbm^k + \ubm^{k-1})$
\State $\ubm^k \leftarrow \ubm^{k-1} + (\zbm^k - \xbm^k)$
\EndFor
\State \Return{$x^t$}
\end{algorithmic}
\end{algorithm}%
\end{minipage}
\hspace{0.25em}
\begin{minipage}[t]{.49\textwidth}
\begin{algorithm}[H]
\caption{PnP-FISTA~\cite{Kamilov-2017}}\label{alg:pnpfista}
\begin{algorithmic}[1]
\State \textbf{input: } $\xbm^0 = \sbm^0$, $\gamma > 0$, and $\theta_k \in (0,1]\, {\forall k }$
\For{$k = 1, 2, \dots, t$}
\State $\zbm^k \leftarrow \xbm^{k-1}-\gamma \nabla g(\xbm^{k-1})$
\State $\sbm^k \leftarrow D(\zbm^k)$
\State $\xbm^k \leftarrow (1-\theta_k) \sbm^k + \theta_k\sbm^{k-1} $
\EndFor
\State \Return{$x^t$}
\end{algorithmic}
\end{algorithm}%
\end{minipage}
\caption{Plug-and-play priors (PnP) refers to a family of iterative algorithms that replace the proximal operator $\prox_{\gamma h}: \R^n \rightarrow \R^n$ of the regularizer $h$ (as in Fig.~\ref{fig:minimization}) by a more general denoiser $D: \R^n \rightarrow \R^n$ in line~4. The success of deep learning in image restoration has led to a wide adoption of PnP for exploiting \emph{learned} priors specified through pre-trained deep convolutional neural nets (CNNs), leading to state-of-the-art performance in a variety of applications.}
\end{figure}

\subsection{Plug-and-Play Priors Algorithms}

The first PnP algorithm was PnP-ADMM~\cite{venkatakrishnan2013school}, which is implemented by iterating the steps in Algorithm~\ref{alg:pnpadmm} until convergence. 
The operator $D$ in PnP-ADMM is an image denoiser that approximates a solution to the problem in Eq.~\eqref{Eq:MAPDenoise}. 
While the original formulation of PnP relies on ADMM, PnP can be equally effective when used with other proximal algorithms, such as \emph{primal-dual splitting (PDS)}~\cite{Ono-2017} and \emph{fast iterative shrinkage/thresholding algorithm (FISTA)}~\cite{Kamilov-2017}. Algorithm~\ref{alg:pnpfista} summarizes the steps of PnP-FISTA, which is based on the traditional FISTA summarized in Algorithm~\ref{alg:fista}. 

PnP-ADMM and PnP-FISTA share the important feature of modularity; they explicitly separate the application of the physical models (data-fidelity update in line 3 of both algorithms) from that of the learned models (image denoising in line 4 of both algorithms). 
This observation reveals a key strength of PnP methods: they can be easily customized for different measurement operators by changing the data-fidelity term, thus enabling the use of the same learned CNN over a wide range of applications without retraining. Similarly, PnP methods provide a simple mechanism to combine different image priors on the same problem by simply changing the denoiser $D$. Note that since the prior in PnP is learned on a pretext task (\emph{image denoising}) rather than on the final task (\emph{image reconstruction}), PnP can be considered a \emph{self-supervised learning} framework.

\begin{figure}[t]
\centering\includegraphics[width=15cm]{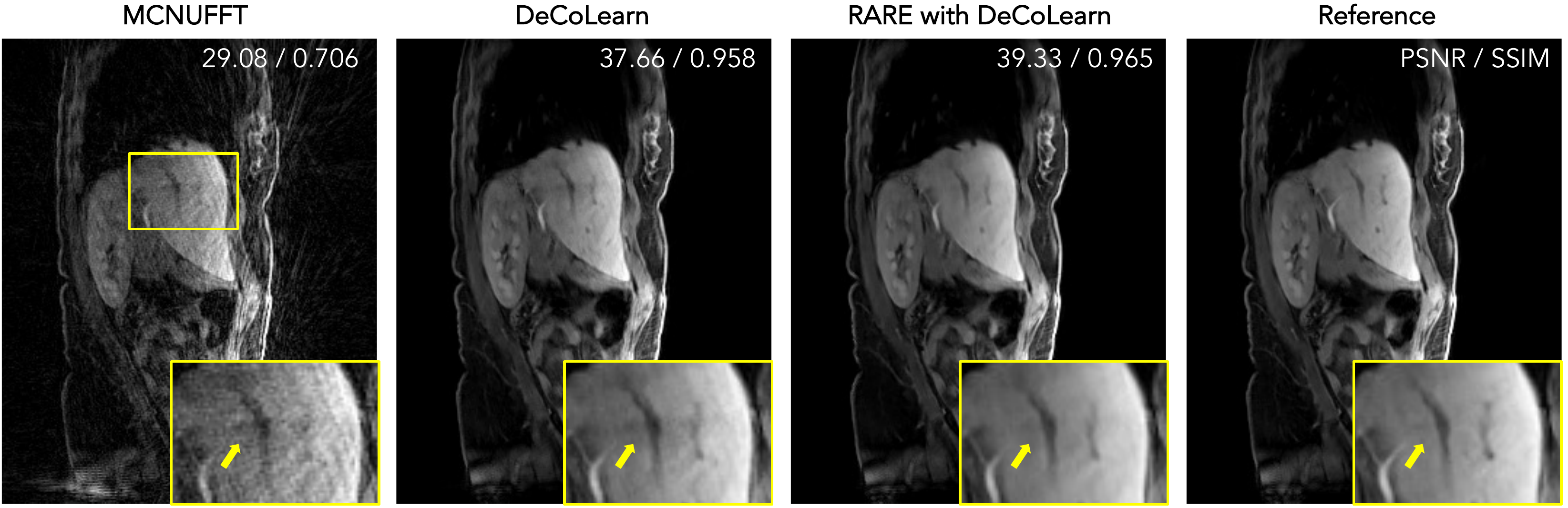}
\caption{PnP algorithms explicitly separate the application of the forward model from that of the learned prior, enabling the adaptation of trained CNNs to new sensor configurations. This is illustrated on experimentally collected 3D MRI data corresponding to 800 radial spokes (scans of about 2 minute). MCNUFFT refers to a simple inversion of the measurement operator without any regularization. DeCoLearn~\cite{Gan.etal2021} is a CNN that was trained under a mismatched sensor configuration corresponding to 1600 lines (scans of about 4 minutes). A variant of PnP called RARE~\cite{Liu.etal2020} is used to adapt DeCoLearn to the desired 800 line data. The results of the DeCoLearn reconstruction using all the available 2000 lines is shown as Reference. The numbers on the top-right corner correspond to the relative PSNR/SSIM values with respect to Reference. Note the ability of RARE to successfully adapt DeCoLearn to 800-line data.}
\label{Fig:RARE}
\end{figure}

One of the practical differences between various PnP algorithms is the treatment of the data-fidelity term, $g$, which models the physical measurements. PnP-FISTA uses a standard (explicit) gradient descent step $\zbm = \xbm - \gamma \nabla g(\xbm)$, while PnP-ADMM uses the proximal operator $\prox_{\gamma g}$, which can be written as an implicit gradient step $\zbm = \prox_{\gamma g}(\xbm) = \xbm - \gamma \nabla g(\zbm)$, with $\nabla g$ evaluated at $\zbm$. We assume for simplicity that $g$ is differentiable -- extensions to nondifferentiable $g$ require more care.

For the loss $g(x) = \frac{1}{2} \|\Abm\xbm - \ybm\|_2^2$, these updates can be computed as
\begin{equation}
\label{Eq:DataGradProx}
\xbm - \gamma \nabla g(\xbm) = \xbm - \Abm^\mathsf{T}(\Abm\xbm-\ybm) \quad\text{and}\quad \prox_{\gamma g}(\xbm) = \left(\bm{I}+\gamma \Abm^\mathsf{T}\Abm\right)^{-1}(\xbm + \gamma \Abm^\mathsf{T}\ybm)\,.
\end{equation}

PnP-ADMM is known for its fast empirical convergence and efficiency for many widely-used operators in computational imaging. 
However, it requires the computation of the proximal map, whereas PnP-FISTA requires only the computation of the  gradient, $\nabla g$.
In principle, the gradient is simpler than the proximal map, but in many applications, the proximal map can be computed or approximated efficiently using general methods such as conjugate gradient, or with specialized methods,
such as when the forward model is a spatial blurring operator that can be computed using the fast Fourier transform (FFT) \cite{Teodoro.etal2019}.
In other cases, the proximal map can be efficiently computed using partial updates \cite{Sridhar2020} that avoid the explicit inversion of ${(\bm{I}+\gamma \Abm^\mathsf{T}\Abm)}$ in Eq.~\eqref{Eq:DataGradProx}.  This is accomplished by maintaining an additional state variable that is used as initialization for the proximal minimization problem.  The minimization is approximated by a few steps of an iterative solver, starting from this initialization.  As the outer loop converges, this additional state variable also converges \cite{Sridhar2020}, so these partial updates reduce computation without compromising the accuracy of the final solution.

An important conceptual point is that PnP algorithms with black-box denoisers do not generally solve an optimization problem. That is, the original ADMM and FISTA algorithms solve the optimization problem in \eqref{eq:RegularizedInversion}.  But once the proximal map denoising operation is replaced with a black-box denoiser, $D$, then there is no longer any corresponding function $h$ to minimize.  In fact, the numerical evaluation of many widely-used denoisers, including BM3D and DnCNN, reveals that their Jacobians are not symmetric, which implies that these denoisers are neither proximal maps nor gradient descent steps \cite[Theorem 1]{Reehorst2019}. 

Nonetheless, it is still possible to formulate a criterion for the converged solution for PnP using a consensus equilibrium formulation \cite{Buzzard.etal2017} (see also Section~\ref{sec:mace} below) given by
\begin{equation}
\label{Eq:FixedPointInterpretation}
\xbm = G(\xbm - \ubm) \quad\text{and}\quad \xbm = D(\xbm + \ubm)\,,
\end{equation}
where $G \defn \prox_{\gamma g}$ and $\xbm$, $\ubm$ are the converged values of PnP-ADMM. Interestingly, in the consensus equilibrium equation of~\eqref{Eq:FixedPointInterpretation}, $\xbm$ is the final reconstruction, and $\ubm$ can be interpreted as noise that is removed by the denoiser in $\xbm = D(\xbm + \ubm)$ on the one hand and balanced by the action of the data fitting update in $\xbm = G(\xbm - \ubm)$ on the other.

\begin{tutorialbox}{Implementing PnP-ADMM Super-resolution in SCICO}

\label{scicoexample}

SCICO \cite{scico-2022} is an open source library for computational imaging that includes implementations of PnP algorithms. Since SCICO is build on top of JAX, it provides seamless support for learned deep priors. Below we show the steps for implementing PnP-ADMM for the example in Fig.~\ref{Fig:SR}.
We assume that a reference image has been loaded as the variable \texttt{img}, then set up problem parameters, such as the downsampling rate and noise level, and construct a downsampled and noise-corrupted measurement that will be superresolved:
\begin{lstlisting}[language=Python]
rate = 4  # downsampling rate
σ = 2e-2  # noise standard deviation
Afn = lambda x: downsample_image(x, rate=rate)  # forward operator 
s = Afn(img)  # downsample reference image
noise, key = scico.random.randn(s.shape, seed=0)
sn = s + σ * noise  # downsampled and noise-corrupted measurement
\end{lstlisting}
We next set up the inverse problem of form~\eqref{eq:RegularizedInversion}, where $g$ is the least-squares function and $h$ is used to invoke DnCNN as the black-box denoiser.
\begin{lstlisting}[language=Python]
A = linop.LinearOperator(input_shape=img.shape, output_shape=s.shape,
                         eval_fn=Afn)
g = loss.SquaredL2Loss(y=sn, A=A)
C = linop.Identity(input_shape=img.shape)
h = functional.DnCNN("17M")
\end{lstlisting}
We obtain a baseline solution (and initializer for PnP) by denoising the pseudo-inverse of the forward operator.
\begin{lstlisting}[language=Python]
xpinv, info = solver.cg(A.T @ A, A.T @ sn, snp.zeros(img.shape))
dncnn = denoiser.DnCNN("17M")  # construct denoiser object
xden = dncnn(xpinv)  # denoised pseudo-inverse solution
\end{lstlisting}
Finally, we set up ADMM to solve the inverse problem.
\begin{lstlisting}[language=Python]
ρ = 3.4e-2  # ADMM penalty parameter
solver = ADMM(f=g, g_list=[h], C_list=[C], rho_list=[ρ], x0=xden, 
              maxiter=12, itstat_options={"display": True},
              subproblem_solver=LinearSubproblemSolver(
                  cg_kwargs={"tol": 1e-3, "maxiter": 10}
              )
)
xppp = solver.solve()  # PnP solution
\end{lstlisting}
\end{tutorialbox}

To establish \eqref{Eq:FixedPointInterpretation}, note that the fixed points $\zbm, \xbm$, and $\ubm$ of the PnP-ADMM iteration satisfy
\begin{equation}
\zbm = G(\xbm-\ubm)\,, \quad \xbm = D(\zbm + \ubm)\,, \quad \ubm = \ubm + \zbm - \xbm\,.
\end{equation}
From the last equation we conclude that $\xbm = \zbm$, which leads directly to Eq.~\eqref{Eq:FixedPointInterpretation}.  Also, the first-order optimality condition for the minimization problem  $\xbm = G(\xbm-\ubm) = \prox_{\gamma g}(\xbm-\ubm)$ is  $ 0 = \xbm - (\xbm - \ubm) + \gamma \nabla g(\xbm)$, so  $\ubm = -\gamma \nabla g(\xbm)$.  
A similar analysis in ~\cite{Meinhardt.etal2017,Sun2019a} shows that the fixed points of PnP-FISTA satisfy the same consensus equilibrium conditions:  
\begin{equation}  \label{eq:FixedPointFISTA}
\xbm = D(\xbm - \gamma \nabla g(\xbm)) 
\quad\Leftrightarrow\quad 
\begin{cases}
\xbm+\ubm = \xbm-\gamma \nabla g(\xbm) \\
\xbm = D(\xbm+\ubm)
\end{cases}
\quad\Leftrightarrow\quad 
\begin{cases}
\xbm = G(\xbm-\ubm) \\
\xbm = D(\xbm+\ubm)\,,
\end{cases}
\end{equation}
where we again used the first-order optimally condition to convert from $\nabla g$ to $G$.

The convergence of PnP algorithms can be established using \emph{monotone operator theory}~\cite{Bauschke2011}. In this approach, as in~\cite{sreehari2016TCI, Chan2017, Buzzard.etal2017, Meinhardt.etal2017, Sun2019a, Teodoro.etal2019, Ryu2019},  the problem is first expressed as finding a fixed point of some high-dimensional operator, which under appropriate hypotheses can be iterated to yield a solution.  

The proof of convergence of PnP-ADMM~\cite{Buzzard.etal2017, Ryu2019} begins by showing that the fixed points of PnP-ADMM are in 1-to-1 correspondence with the fixed points of 
the operator  
\begin{equation}
\label{Eq:ADMMOperator}
T \defn (2G-I)(2D-I)\,.
\end{equation}
In fact, after a linear change of coordinates, Algorithm~\ref{alg:pnpadmm} is equivalent to the Mann iterations of $T$ given by $\vbm^k \leftarrow \frac{1}{2} \vbm^{k-1} + \frac{1}{2} T(\vbm^{k-1})$~\cite{Parikh.Boyd2014, Buzzard.etal2017}. This yields linear convergence to a unique fixed point when $T$ is a contraction, which is true when $g$ is strongly convex and $R \defn I - D$ is a sufficiently strong contraction~\cite{Ryu2019}.  Weaker conditions establish sublinear convergence to a possibly non-unique fixed point~\cite{Sun.etal2021}.
Other well-known theoretical results on PnP-ADMM include its convergence for implicit proximal operators~\cite{sreehari2016TCI}, for bounded denoisers~\cite{Chan2017}, and for linearized Gaussian mixture model (GMM) denoisers~\cite{Teodoro.etal2019}. 
Even CNN-based denoisers can be trained to satisfy these contractive, nonexpansive, or Lipschitz conditions 
by 
using \emph{spectral normalization} techniques~\cite{Ryu2019, Sun.etal2019b}.

The convergence of PnP-ISTA (which is PnP-FISTA with the Nesterov acceleration parameters set to $\theta_k = 1$ for all $k \geq 1$) can be established by expressing it using operator $F \defn D(I-\gamma \nabla g)$. When the data-fidelity term $g$ is strongly convex and the denoiser $D$ is Lipschitz continuous with a sufficiently small constant, then the iteration of $F$ converges linearly to its unique fixed point~\cite{Ryu2019}. On the other hand, when $g$ is weakly convex and the denoiser $D$ is firmly nonexpansive, then the iteration converges sublinearly to its fixed point~\cite{Sun2019a}. Related results have shown that PnP-ISTA converges to a minimizer of some global cost function when the denoiser $D$ corresponds to a \emph{minimum mean squared error (MMSE)} estimator applied to the denoising problem in Eq.~\eqref{Eq:MAPDenoise}~\cite{Xu.etal2020}, and have established recovery guarantees for PnP for the measurement operators that satisfy the \emph{restricted eigenvalue condition (REC)} commonly used in compressive sensing~\cite{Liu.etal2021b}.

\subsection{Regularization by Denoising}

\emph{Regularization by denoising (RED)} is an algorithm inspired by PnP that also enables integration of denoisers as priors for inverse problems~\cite{Romano2017}. The \emph{steepest descent} variant of RED (RED-SD)~\cite{Romano2017} can be implemented by iterating the following steps until convergence
\begin{equation}
\label{Eq:REDSD}
\xbm^k \leftarrow \xbm^{k-1} - \gamma H(\xbm^{k-1}) \quad\text{with}\quad H(\xbm) \defn \nabla g(\xbm) + \tau (\xbm - D(\xbm))\,,
\end{equation}
where $\gamma > 0$ is the step size, $D$ is a denoiser, and $\tau > 0$ is the regularization parameter. 

While RED was initially derived as an optimization problem~\cite{Romano2017}, a subsequent analysis~\cite{Reehorst2019} showed that an interpretation as a fixed point problem is more appropriate for practical denoisers.
Using this approach, the fixed-point condition in \eqref{Eq:REDSD} translates to the RED equilibrium condition given by
\begin{equation} \label{eq:RED-equilibrium}
 -\nabla g(\xbm) = \tau (\xbm - D(\xbm)) \ .   
\end{equation}
As noted in Eq.~\eqref{eq:gh}, the function $g$ is often the negative log-likelihood of the distribution $p_{\ybm|\xbm}$ relating the reconstruction $\xbm$ to the measurements $\ybm$.  In this setting $-\nabla g$ is known as the ``score'' of this distribution. This negative gradient describes the steepness of the log-likelihood function and hence the sensitivity to changes in $\xbm$.  From this, we see that Eq. \eqref{eq:RED-equilibrium} balances changes in the log-likelihood against the update step $\xbm - D(\xbm)$.  This is similar to the balance in Eq.~\eqref{Eq:FixedPointInterpretation}, in which the same $\ubm$ is removed by the denoiser and added by the data-fitting map.  A much more complete discussion of RED algorithms and score matching is given in \cite{Reehorst2019}.

The convergence of RED algorithms can also be analyzed using monotone operator theory. In particular, it can be shown that for a convex function $g$ and a nonexpansive denoiser $D$, RED-SD converges sublinearly to a set of $\xbm$ satisfying the equilibrium condition in \eqref{eq:RED-equilibrium}~\cite{Reehorst2019, Sun.etal2019b}.

\begin{figure}[t]
\centering\includegraphics[width=15cm]{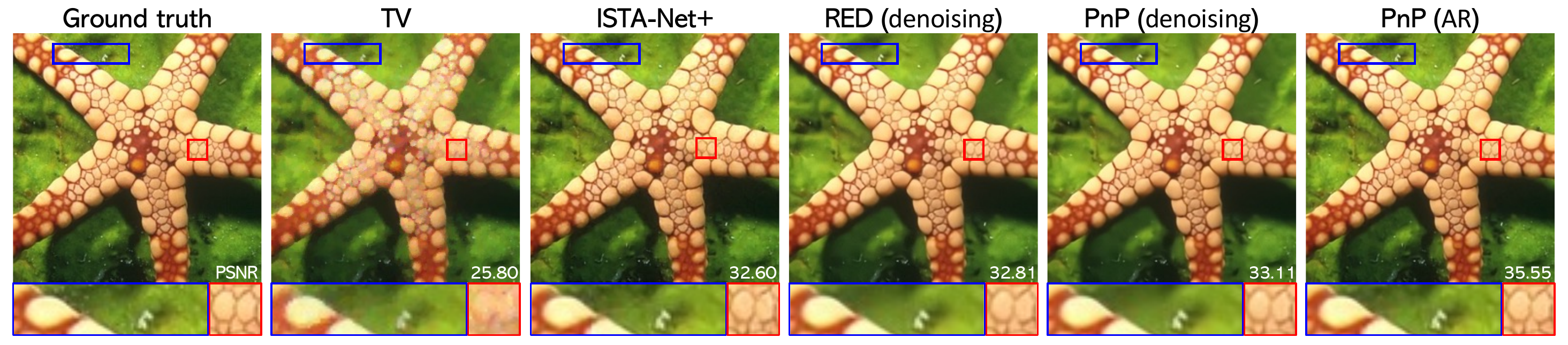}
\caption{Visual evaluation of color image recovery in compressive sensing from random projections with 20\% subsampling. The results of total variation (TV) and a well-known deep unfolding (DU) architecture ISTA-Net+ are provided for reference. The methods PnP (denoising) and RED (denoising) use a pre-trained AWGN denoiser as an image prior. The method PnP (AR) uses a problem-dependent artifact-removal (AR) operator pre-trained using DU. Note that the choice of denoiser affects the reconstruction significantly (PSNR shown in white).}
\label{Fig:PnPREDComparison}
\end{figure}

Fig.~\ref{Fig:PnPREDComparison} presents results using PnP and RED on compressive sensing from random projections with 20\% subsampling. The setup used in the simulation is identical to that described in~\cite{Liu.etal2021b}. The results of the traditional TV reconstruction and of the ISTA-Net+ deep unfolding (DU) architecture~\cite{Zhang.Ghanem2018} are presented for reference (see Section~\ref{sec:dudeq} for the discussion on DU). The figure considers two priors for PnP: (a) an AWGN denoiser and (b) an artifact-removal (AR) operator trained to remove artifacts specific to the PnP iterations that is used in place of an AWGN denoiser. Both priors are implemented using the DnCNN architecture, with its batch normalization layers removed to enable control of the Lipschitz constant of the network via spectral normalization. The AR operator $D$ is trained by including it into a DU architecture that performs PnP iterations and training it end-to-end in a supervised fashion.  This approach has the disadvantage that the prior model is no longer completely separate from the forward model, but as seen in Fig.~\ref{Fig:PnPREDComparison}, it can yield significantly improved results relative to an AWGN denoiser.  

\subsection{Online Plug-and-Play Algorithms}

\begin{figure}
\linespread{1}
\begin{minipage}[t]{.49\textwidth}
\begin{algorithm}[H]
\caption{Online PnP~\cite{Sun2019a}}\label{alg:onppp}
\begin{algorithmic}[1]
\State \textbf{input: } $\xbm^0$, $b \geq 1$, and $\gamma > 0$
\For{$k = 1, 2, \dots, t$}
\State Choose an index $i_k \in \{1, \dots, b\}$
\State $\zbm^k \leftarrow \xbm^{k-1}-\gamma \nabla g_{i_k}(\xbm^{k-1})$
\State $\xbm^k \leftarrow D(\zbm^k)$
\EndFor
\end{algorithmic}
\end{algorithm}%
\end{minipage}
\hspace{0.25em}
\begin{minipage}[t]{.49\textwidth}
\begin{algorithm}[H]
\caption{SIMBA~\cite{Wu.etal2020}}\label{alg:onred}
\begin{algorithmic}[1]
\State \textbf{input: } $\xbm^0$, $b \geq 1$, $\gamma > 0$, and $\tau > 0$
\For{$k = 1, 2, \dots, t$}
\State Choose an index $i_k \in \{1, \dots, b\}$
\State $H_{i_k}(\xbm^{t-1}) \leftarrow \nabla g_{i_k}(\xbm^{t-1}) + \tau R(\xbm^{t-1})$
\State $\xbm^k \leftarrow \xbm^{k-1}-\gamma H_{i_k}(\xbm^{k-1})$
\EndFor
\end{algorithmic}
\end{algorithm}%
\end{minipage}
\caption{The complexity of evaluating the batch gradient of $\nabla g$ or proximal operator $\prox_{\gamma g}$ is computationally expensive in some applications. This has motivated the development of online, stochastic, or incremental variants of PnP that use a single element or a small subset of the measurements at each iteration. The per-iteration complexity of such algorithms is independent of the batch size $b \geq 1$, thus making them suitable for certain large-scale applications.}
\end{figure}

\begin{figure}[t]
\centering\includegraphics[width=15cm]{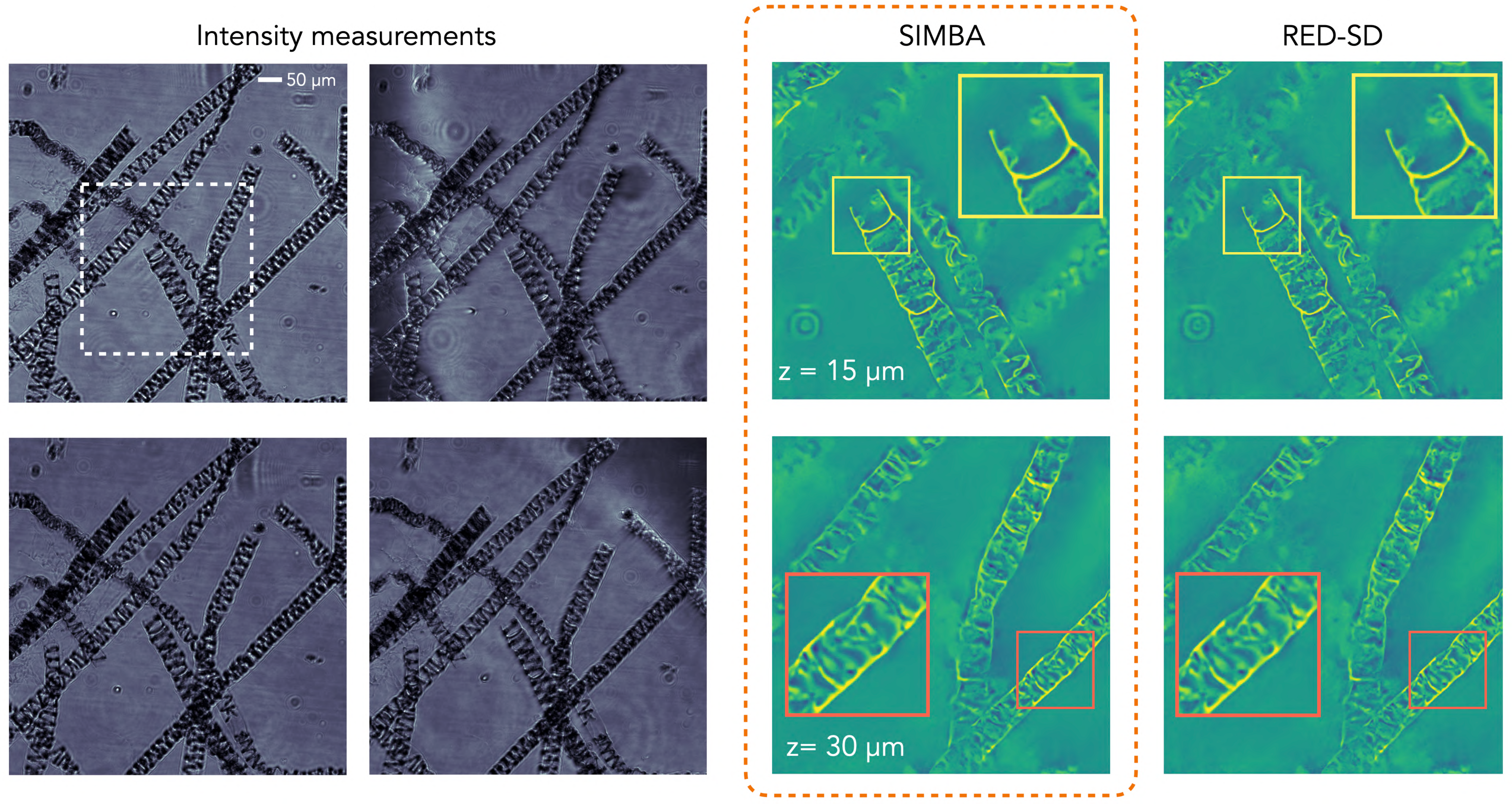}
\caption{Online PnP algorithms, such as SIMBA in Algorithm~\ref{alg:onred}, can reduce the computational and memory complexity of PnP. Here, we show the reconstruction of a 3D algae sample from $89$ experimentally collected \emph{intensity diffraction tomography (IDT)} measurements (see four images on the left). SIMBA, which uses minibatches of size $p = 10$, is compared against RED-SD, which uses all $b = 89$ measurements at each iteration. Both algorithms use exactly the same measurement model and the same DnCNN AWGN denoiser. Note how the results of SIMBA are indistinguishable from RED-SD even though the per-iteration complexity of SIMBA is only a fraction of that of RED-SD.}
\label{Fig:SIMBA}
\end{figure}

The traditional PnP methods are \emph{batch} algorithms in the sense that they compute the gradient $\nabla g$ or the proximal operator $\prox_{\gamma g}$ of the data-fidelity term $g$ by using the whole measurement vector $\ybm \in \R^m$. The per-iteration computational and memory complexity of batch PnP algorithms depends on the total number of measurements. For example, in tomography with $b$ projections the complexity of evaluating $\nabla g$ scales linearly with $b$, making it computationally expensive for large $b$. This has motivated interest in \emph{online}, \emph{stochastic}, and/or \emph{incremental} PnP algorithms that approximate the batch $\nabla g$ with an approximation $\widehat{\nabla} g$ based on a single element or a small subset of the measurements~\cite{Sun2019a, Sun.etal2021, Wu.etal2020}. Consider the decomposition of $\R^m$ into $b \geq 1$ blocks
\begin{equation}
\R^m = \R^{m_1} \times \R^{m_2} \times \cdots \times \R^{m_b} \quad\text{with}\quad m = m_1 + m_2 + \cdots + m_b\,.
\end{equation}
In this setting, the data-fidelity term and the corresponding gradient vector are given by
\begin{equation}
g(\xbm) = \frac{1}{b}\sum_{i = 1}^b g_i(\xbm) \quad\text{and}\quad \nabla g(\xbm) =  \frac{1}{b}\sum_{i = 1}^b \nabla g_i(\xbm)\,,
\end{equation}
where each $g_i$ is evaluated only on the subset $\ybm_i \in \R^{m_i}$ of the full measurement vector $\ybm \in \R^m$. For example, each individual term in the gradient can be set to the quadratic function ${g_i(\xbm) = \frac{1}{2}\|\ybm_i - \Abm_i\xbm\|_2^2}$, where $\Abm_i$ is the operator corresponding to the measurement block $\ybm_i$. 

\begin{figure}[t]
\centering\includegraphics[width=15cm]{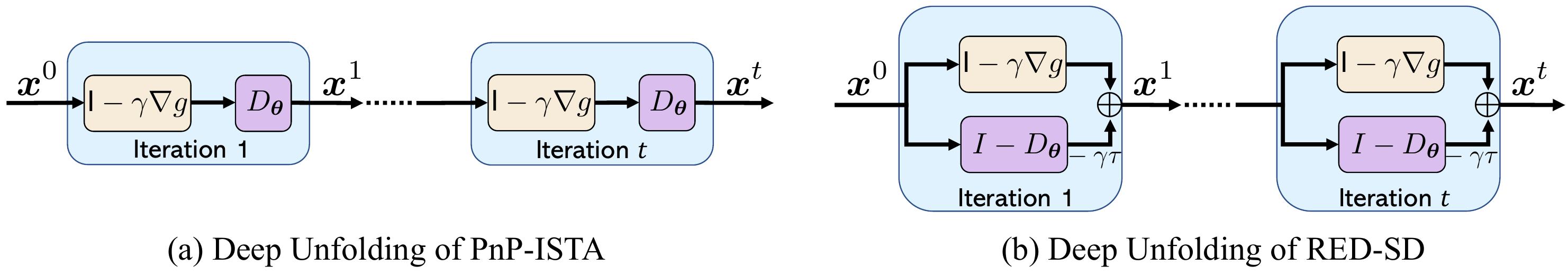}
\caption{The PnP framework is related to two other popular computational-imaging paradigms, deep unfolding (DU) and deep equilibrium models (DEQ). A PnP algorithm, such as PnP-ISTA or RED-SD, can be turned into a DU architecture by truncating the algorithm to $t \geq 1$ iterations and training the weights $\bm{\theta}$ of the CNN $D_{\bm{\theta}}$ end-to-end. Similarly, a DEQ architecture can be obtained by running the PnP algorithm until convergence and using the implicit differentiation at the fixed point to train the weights $\thetabm$. The operator $D_{\bm{\theta}}$ in DU/DEQ is not necessarily an AWGN denoiser, instead it is an artifact-removal (AR) operator trained to remove artifacts specific to the PnP iterations.}
\label{Fig:DeepUnfolding}
\end{figure}

Online PnP in Algorithm~\ref{alg:onppp} and SIMBA in Algorithm~\ref{alg:onred} are online extensions of PnP-FISTA and RED-SD, respectively. Both algorithms improve the scalability to large-scale measurements by using only a single component gradient $\nabla g_{i_k}(\xbm)$, with $i_k \in \{1, \dots, b\}$, making their per-iteration complexity independent of $b$. Online PnP algorithms can be implemented using different block selection rules. The strategy commonly adopted for the theoretical analysis focuses on selecting indices $i_k$ as independent and identically distributed (i.i.d.) random variables distributed uniformly over $\{1, \dots, b\}$. An alternative would be to proceed in epochs of $b$ consecutive iterations, where the set $\{1, \dots, b\}$ is reshuffled at the start of each epoch, and the index $i_k$ is selected from this ordered set. Online PnP algorithms can also be implemented in a \emph{minibatch} fashion by replacing $\nabla g_{i_k}$ in Step 4 of both by a minibatch gradient
\begin{equation}
\label{Eq:MinibatchGradient}
\widehat{\nabla} g(\xbm) = \frac{1}{p} \sum_{j = 1}^p \nabla g_{i_j} (\xbm) \,,
\end{equation}
where $p$ is the minibatch size and  $i_1, \dots, i_p$ are indices selected from the set $\{1, \dots, b\}$. The minibatch variants of online PnP can process several blocks at parallel in every iteration, thus improving efficiency on multi-processor hardware architectures. While online algorithms have traditionally focused on partial approximations of the gradient, recent work has also explored the approximation of the batch proximal operator $\prox_{\gamma g}$ in PnP-ADMM by a partial proximal operator $\prox_{\gamma g_i}$~\cite{Sun.etal2021}. One can also consider block-coordinate extensions on online PnP by considering decomposition of the image space $\R^n$ into a number of smaller image blocks~\cite{Sun.etal2019b}. 

The fixed-point convergence analysis of online PnP algorithms uses mathematical tools from traditional stochastic optimization and monotone operator theory. The key requirement for the analysis is that the gradient estimate in Eq.~\eqref{Eq:MinibatchGradient} is unbiased and has bounded variance
\begin{equation}
\E\left[\widehat{\nabla} g(\xbm)\right] = \nabla g(\xbm) \quad\text{and}\quad \E\left[\|\nabla g(\xbm) - \widehat{\nabla} g(\xbm)\|_2^2\right] \leq \frac{\nu^2}{p}\,,
\end{equation}
for some constant $\nu > 0$ and every $\xbm \in \R^n$. Note that when $i_k$ is selected uniformly at random from $\{1,\dots,b\}$, the unbiasedness assumption is automatically satisfied. Then, for convex data-fidelity terms $g_i$ and firmly nonexpansive denoisers $D$, one can establish the sublinear convergence of online PnP algorithms to their fixed points~\cite{Sun2019a, Sun.etal2021, Wu.etal2020}.

\subsection{Deep Unfolding and Deep Equilibrium Models}
\label{sec:dudeq}

\emph{Deep unfolding (DU)} (also known as \emph{deep unrolling} or \emph{algorithm unrolling}) is a DL paradigm with roots in sparse coding \cite{r2-gregor2010learning,r3-chen_theoretical_2018}
that has gained popularity in computational imaging due to its ability to provide a systematic connection between iterative algorithms and deep neural network architectures~\cite{r3-chen_theoretical_2018,Monga.etal2021}. Many PnP algorithms have been turned into DU architectures by parameterizing the operator $D_{\thetabm}$ as a CNN with weights $\thetabm$, truncating the PnP algorithm to a fixed number of iterations $t \geq 1$, and training the corresponding architecture end-to-end in a supervised fashion. For example, Fig.~\ref{Fig:DeepUnfolding} illustrates the representation of $t$ iterations of PnP-ISTA and RED-SD as DU architectures.

Consider a set of paired data $(\xbm_i, \ybm_i)$, where $\xbm_i$ is the desired ``ground truth'' image and $\ybm_i = \Abm\xbm_i+\ebm_i$ is its noisy observation. Consider also the iterate $\xbm_i^t(\thetabm)$ of a PnP algorithm truncated to $t \geq 1$ iterations, where we made explicit the dependence of the PnP output on the weights $\thetabm$ of the CNN parameterizing $D_{\thetabm}$. DU interprets the steps required for mapping the input vector $\ybm_i$ and the initialization $\xbm_i^0$ to the output $\xbm_i^t(\thetabm)$ as layers of a deep neural network architecture. The DU training is performed by solving the optimization problem
\begin{equation}
\label{Eq:DUTraining}
\widehat{\thetabm} = \argmin_{\thetabm} \sum_{i} L(\xbm_i, \xbm_i^t(\thetabm))\,,
\end{equation}
where $L$ is a loss function that quantifies the discrepancy between the true and predicted solutions. Once trained using Eq.~\eqref{Eq:DUTraining}, the truncated PnP algorithm can be directly used for imaging~\cite{Liu.etal2021b}.

\emph{Deep equilibrium models (DEQ)} is a recent extension of DU to an arbitrary number of iterations~\cite{Gilton.etal2021}. DEQ can be implemented by replacing $\xbm_i^t(\thetabm)$ in Eq.~\eqref{Eq:DUTraining} by a fixed-point $\overline{\xbm}_i(\thetabm)$ of a given PnP algorithm and using implicit differentiation for updating the weights $\thetabm$. The benefit of DEQ over DU is that it doesn't require the storage of the intermediate variables for solving Eq.~\eqref{Eq:DUTraining}, thus reducing the memory complexity of training. However, DEQ requires the computation of the fixed-point $\overline{\xbm}_i(\thetabm)$, which can increase the computational complexity.

There are some important differences between traditional PnP and DU/DEQ. Traditional PnP relies on an AWGN denoiser as an image prior. On the other hand, the operator $D_{\thetabm}$ in DU/DEQ is not an AWGN denoiser; instead, it is an artifact-removal (AR) operator trained to remove artifacts specific to the PnP iterations. As seen in Fig.~\ref{Fig:PnPREDComparison}, which shows the relative performance of PnP using an AWGN denoiser and using a pre-trained AR operator, this problem-specific training can yield significantly improved results.  However, this performance comes at a cost;  while the prior in traditional PnP is \emph{fully decoupled} from the measurement operator, that of DU/DEQ is trained by accounting for the measurement operator $\Abm$.  Hence the DU/DEQ approach has reduced generality and higher computational/memory complexity of training, since the AR prior is trained for the specific task of reconstruction from random projections rather than for AWGN denoising. 

\subsection{Related Approaches}
\label{sec:related}

There are a wide variety of approaches to learning and using prior information in the context of inverse imaging, far too many to describe completely.  Early work on non-CNN learned priors includes EPLL \cite{r4-zoran2011learning}, which uses a cost-function approach on patches.  Work related to the use of CNNs as priors includes \cite{r5-zhang2017learning}, which describes certain empirical advantages enjoyed by CNN denoisers; \cite{r6-tirer2019image}, which uses CNN denoisers with (F)ISTA on a modified cost function with convergence/accuracy benefits \cite{r7-tirer2020back}; and \cite{r8-arjomand_bigdeli_deep_2017}, which is related to RED with a motivation that comes from an analysis of denoising auto-encoders \cite{r9-alain2014what}.

In addition to DU/DEQ, another approach to improving the performance of PnP-inspired methods is to fine-tune denoisers for a specialized distribution of images.  Examples of this include \cite{r10-teodoro2016image}, with images from the same class as the observed image, \cite{r11-teodoro2019convergent}, with images from the same scene as the observed image, and \cite{r12-tirer2019super}, with training on the single observed image.  

\subsection{Trade-offs and Limitations}
\label{sec:limitations}

A key idea of PnP-inspired methods is to encapsulate Bayesian prior information into an algorithmic denoiser.  This approach has the benefit of promoting code modularity in that data-fitting updates and denoisers can be developed separately, with many possible pairings of data updates and denoisers. The down side of this generality is that some reconstruction quality is lost; under ideal conditions, an end-to-end trained system can outperform a general-purpose PnP system.  DU and DEQ methods fall somewhere in the middle in that they have separate data-update and denoising modules, but the denoiser is trained as part of an end-to-end system to enhance reconstruction quality.  

We note also that as with any inversion method, particularly one with learned priors, PnP methods involve a number of hyperparameters to be tuned.  For PnP-ADMM and PnP-FISTA, one of the most important of these is the distribution of images used to train a denoiser, most especially the assumed noise level.   In practice, the noise level needed for optimal reconstruction may not be known at training time.  Approaches to address this include training a denoiser for multiple noise levels \cite{gnanasambandam-2020-one} and reconciling multiple denoisers using MACE  \cite{Buzzard.etal2017}.  An additional factor is the architecture of the neural network, which can play an important role in the quality of results and in the time required for reconstructions.  However, this is a complex design problem with much on-going work and many problem-specific considerations.

Finally, the use of learned priors introduces the possibility of mismatch between training data and application data.  Some work on the effect of such mismatch is described in \cite{r12-tirer2019super,r13-Shocher_2018_CVPR, r14-9156619}. 

\section{Multi-Agent Consensus Equilibrium} \label{sec:mace}

\emph{Multi-Agent Consensus Equilibrium (MACE)}, introduced in \cite{Buzzard.etal2017}, is a framework that extends PnP-ADMM to more than 2 update terms and that provides an equilibrium interpretation to the problem solved by PnP-ADMM.  As noted above, PnP-ADMM with a black-box denoiser does not generally solve an optimization problem, but instead solves the equilibrium problem in Eq.~\eqref{Eq:FixedPointInterpretation}.  MACE takes this equilibrium condition as a starting point and extends it to allow for multiple types of models, including physics-based, data-driven, or application-specific models.  When a physics-based forward model and a denoiser-based prior model are used, then the MACE solution is the same as the PnP solution.  But MACE is more general and offers more flexibility in the choice of models as well as algorithms for computing the solution.

A MACE agent is a function $F: \R^n \rightarrow \R^n$ that takes in an image $\bm{v}$ and produces an ``improved'' output image, $\bm{x} = F(\bm{v})$.  So a denoiser or proximal map are examples of agents, but other agents might implement artifact removal or other heuristic improvements.
We denote the agents by $F_1, \ldots, F_\ell$, each of which maintains its own version $\bm{v}_j$ of the input image.  We can then stack input images as $\bm{v} = [\bm{v}_1, \ldots, \bm{v}_\ell]$ and the output images as $\bm{F}(\bm{v}) = [F_1(\bm{v}_1), \ldots, F_\ell(\bm{v}_\ell)]$. At the same time, we define an averaging operator $\bm{G}(\bm{v}) = (\overline{\bm{v}}, \ldots, \overline{\bm{v}})$, where $\overline{\bm{v}}$ is the average of the $\{\bm{v}_j\}$.  With this notation, the MACE equation is 
\begin{equation}\label{eqn:FG}
\bm{F}(\bm{v}^*) = \bm{G}(\bm{v}^*)\,,
\end{equation}
and the final reconstruction is the average of the vectors $\bm{v}_j^*$.  

\begin{figure}[t]
\centering\includegraphics[width=14cm]{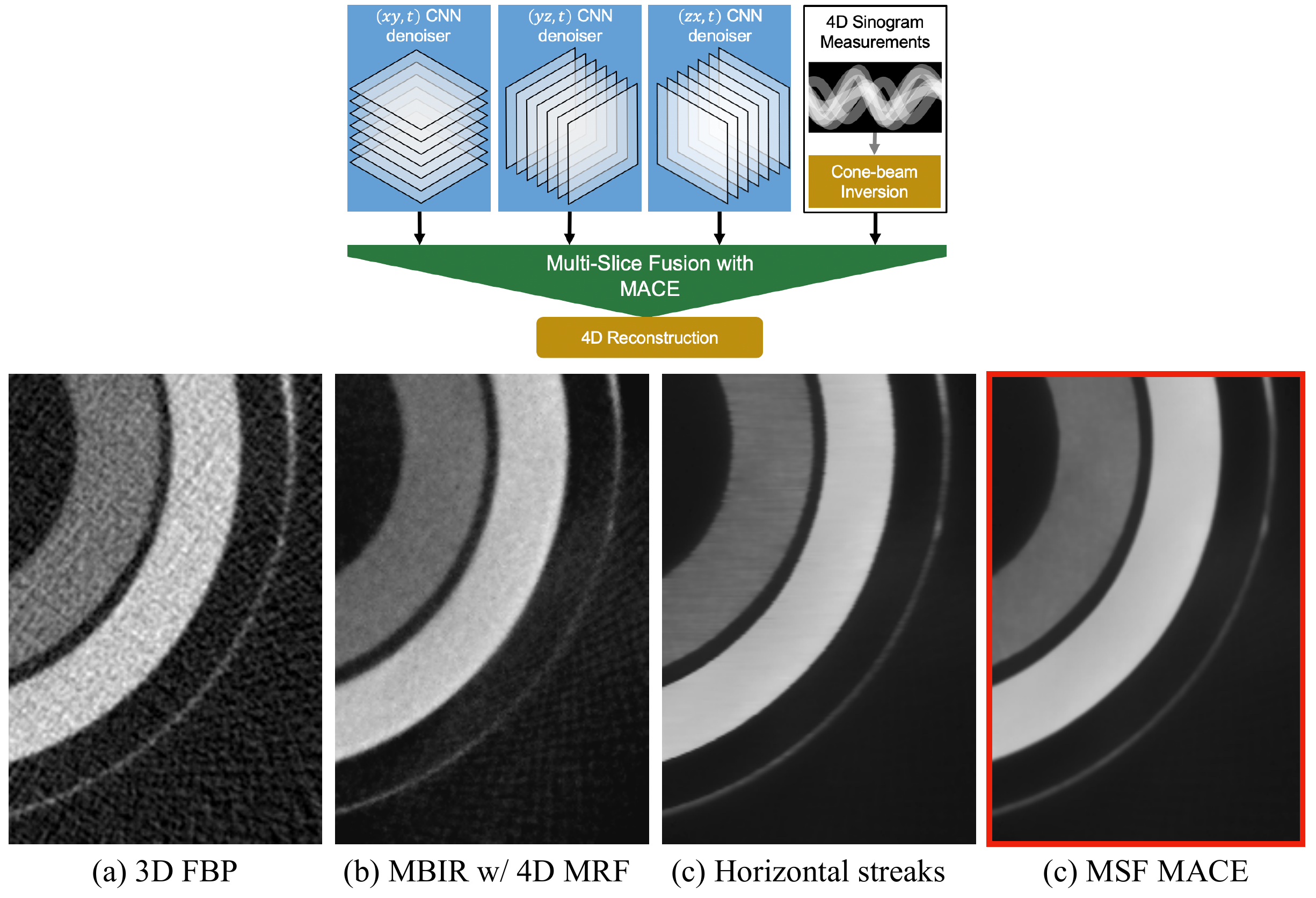}
\caption{Application of MACE from~\cite{majee2021} using multiple 2D denoisers to regularize a time-varying volume (4D).  Top:  Combination of 2D denoisers in multiple orientations with CT measurements using MACE.  Bottom: (a) and (b) show reference reconstructions using FBP and MBIR. (c) shows the reconstruction using 2D denoisers in 2 out of 3 possible spatial orientations, which leads to streaking artifacts in the complementary direction.  (d) shows the Multi-Slice Fusion (MSF) MACE reconstruction using 2D denoisers in all 3 orientations.}
\label{Fig:MSF}
\end{figure}

As seen in the inset ``{\it Geometric Intuition for MACE},'' the MACE equation in \eqref{eqn:FG} has an interpretation as consensus equilibrium. Since each entry in $\bm{G}$ is identical, all agents must output the same reconstruction, so $F_j(\vbm_j) = F_k(\vbm_k)$ for all $j, k$: this is \emph{consensus}.  Moreover, since this consensus point is the same as the average of the input points encoded in $\vbm^\ast$, the updates $\vbm_j - F_j(\vbm_j)$ must sum to 0: this is \emph{equilibrium}.  
As with PnP and RED, the MACE equations can be converted to a fixed point problem by noting that the averaging operator $\bm{G}$ is a linear projection, so has the property that $\bm{G}^2 = \bm{G}$.  This implies that $(2 \bm{G} - \bm{I})$ is its own inverse, which means that $(2 \bm{G} - \bm{I}) (2 \bm{G} - \bm{I}) = \bm{I}$.  Then from Eq.~\eqref{eqn:FG} we see that $2\bm{F}(\bm{v}^*) - \bm{v}^* = 2 \bm{G}(\bm{v}^*) - \bm{v}^*$, so multiplying both sides by $(2 \bm{G} - \bm{I})$ we have that Eq.~\eqref{eqn:FG} is equivalent to 
$$\bm{T}(\bm{v}^*) = \bm{v}^*, \quad\text{where}\quad \bm{T} \defn (2 \bm{G} - \bm{I}) (2 \bm{F} - \bm{I})\,.$$

As above, we can solve this fixed point problem via Mann iterations~\cite{Bauschke2011}. For MACE this takes the form of $\bm{v} \leftarrow (1-\rho) \bm{v} + \rho \bm{T} (\bm{v})$, with $\rho \in (0,1)$.  This is guaranteed to converge when $\bm{T}$ is nonexpansive and has a fixed point, but converges in practice for a wide variety of agents.  The inset \emph{``Geometric Intuition for MACE''} gives additional intuition and a pseudocode implementation of this algorithm for solving the MACE equations.

\begin{tutorialbox}{Geometric Intuition for MACE}

\begin{figure}[H]
\centering
\includegraphics[width=.95\textwidth]{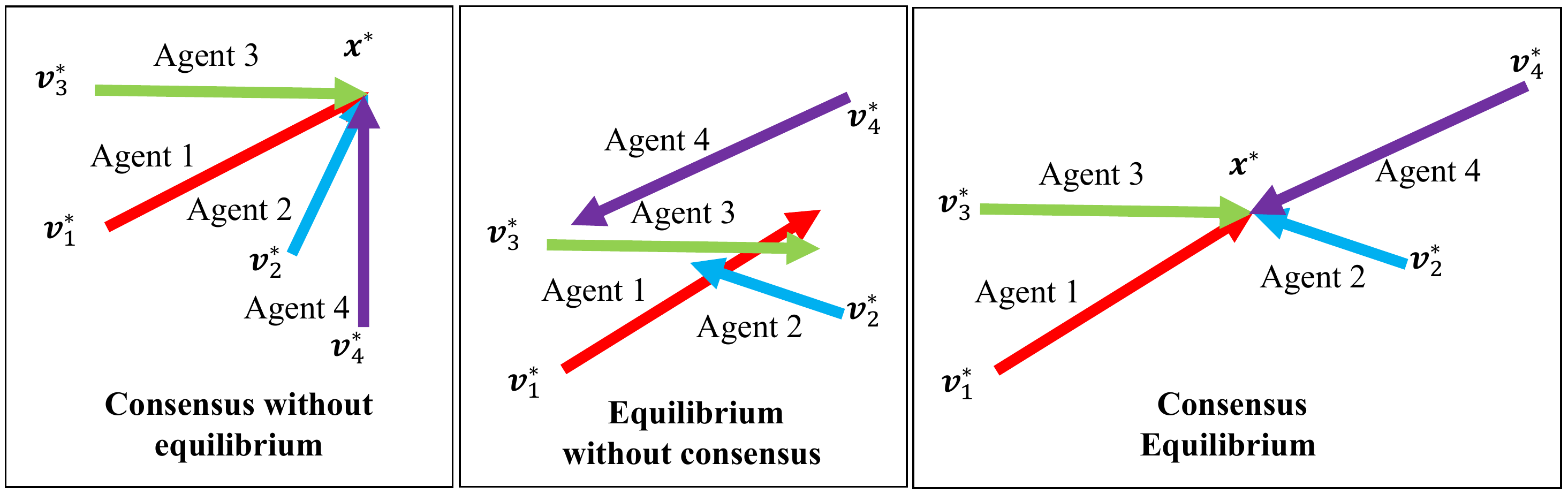}
\caption{Visualization of Multi-Agent Consensus Equilibrium. Consensus means that each agent has the same output, while equilibrium means that vector sum of the updates is $\bm{0}$. }
\label{Fig:MACE}
\end{figure}%
\vspace{-10pt}
\hspace{10pt} MACE is a formal mechanism to compute a reconstruction that is a balance among multiple competing models or agents.  When the agents are a forward and prior model, the MACE solution is exactly the PnP solution. But MACE also works with multiple models, each promoting different desired outcomes.  

\hspace{10pt} An agent is a function, $\bm{x}=F(\bm{v})$, that takes an input image, $\bm{v}$, and makes it better in some way to produce a new image, $\bm{x}$. Denoisers, gradient descent steps and proximal maps are all examples of useful agents.  Beyond that, agents can be neural networks trained to remove application-specific artifacts or noise or other image enhancing operations.  

\hspace{10pt} Algorithm~\ref{alg:mace-mann} shows how to compute the MACE solution with $n$ agents shown as $F_1, \ldots, F_n$.  At convergence, all agents yield the same reconstruction: $F_n(\bm{v_n}) = \bm{x}^*$ for all $n$.  This is the idea of consensus.   This reconstruction is also the average of the input points: $\text{Average}(\bm{v}_1, \ldots, \bm{v}_n) = \bm{v}^*$.  This is the idea of equilibrium.

\begin{algorithm}[H]
\caption{MACE reconstruction~\cite{Buzzard.etal2017}}\label{alg:mace-mann}
\begin{algorithmic}[1]
\State \textbf{input: } Initial Reconstruction $\bm{x}$ (an image) \vspace{-6pt}
\State  $\bm{v} \leftarrow \begin{pmatrix} \bm{x}, \cdots, \bm{x} \\ \end{pmatrix}$ \hspace{1em} \textit{\# Form a stack of images}\vspace{-6pt}
\While{not converged}\vspace{-5pt}
\State $\bm{x} \leftarrow (F_1(\bm{v}_1), \ldots, F_n(\bm{v}_n))$  \hspace{1em}\textit{\# Apply each agent}\vspace{-5pt}
\State $\bm{w} \leftarrow 2\bm{x} - \bm{v}$\vspace{-5pt}
\State $\overline{\bm{z}} \leftarrow$ Average($\bm{w}_1, \ldots, \bm{w}_n$)\vspace{-5pt}
\State $\bm{z} \leftarrow (\overline{\bm{z}}, \ldots, \overline{\bm{z}})$ \hspace{1em} \textit{\# Restack the average image}\vspace{-5pt}
\State $\bm{v} \leftarrow \bm{v}+2\rho (\bm{z}-\bm{x})$  \hspace{1em} \textit{\# Update agent input}\vspace{-5pt}
\EndWhile\vspace{-5pt}
\State $\bm{x}^* \leftarrow \overline{\bm{z}}$ \hspace{1em} \textit{\# Get final result}
\end{algorithmic}
\end{algorithm}%

\end{tutorialbox}

Figure~\ref{Fig:MSF} illustrates the benefits of MACE for fusing the outputs of 3 separate 2D image denoisers to regularize the result of a 4D reconstruction problem in space and time~\cite{majee2021}. 
Each denoiser operates along only 2 of the 3 dimensions, and MACE integrates these 3 denoisers along with a physical model of tomographic projections. 
The image using all 4 agents, labeled \emph{multi-slice fusion (MSF)}, has the best quality, while the images reconstructed with one missing denoising agent contain streaking artifacts aligned with the orientation of the missing agent. 

The use of multiple 2D denoisers has a number of important advantages over 3D/4D denoising. First, 2D processing is more efficient since memory access is more local and the number of nearest neighbors is smaller. Second, 2D denoisers implemented as deep neural networks can be trained on widely available 2D images, whereas 3D/4D data is currently very limited.

\section{PnP in Practice} \label{sec:practice}

\subsection{Implementing PnP algorithms} \label{sec:implementation}

While PnP methods are generally easy to implement, there are several reference implementations that can be used as sources of inspiration and as utilities.  
Some open source libraries that include implementations of PnP algorithms include SPORCO\footnote{\url{https://github.com/bwohlberg/sporco}}, PnP-MACE\footnote{\url{https://github.com/gbuzzard/PnP-MACE}}, and most recently SCICO\footnote{\url{https://github.com/lanl/scico}}, which provides a wide array of computational imaging tools in Python. SCICO is built on JAX, which provides support for seamless code transition between CPU and GPU, acceleration via just-in-time compilation, and automatic differentiation. In particular, the super-resolution demo in \emph{``Turning an Image Denoising Network into Image Super-Resolver''} was implemented using SCICO (see \emph{``Implementing PnP-ADMM in SCICO''}).

\subsection{Applications of PnP} \label{sec:applications}

PnP has been applied to a very wide range of problems, including superresolution \cite{Reid2022} and blind deconvolution, various forms of tomographic imaging \cite{majee2021}, magnetic resonance imaging (MRI) \cite{Liu.etal2020}, and synthetic aperture radar \cite{Pellizzari2020}, to name but a few. In this section, we present two applications of PnP in computational imaging: MRI and \emph{intensity diffraction tomography (IDT)}.

Fig.~\ref{Fig:RARE} presents an application of PnP to a free-breathing 3D MRI problem described in~\cite{Liu.etal2020}. The experimentally collected \emph{in vivo} k-space measurements were acquired using T1-weighted stack-of-starts 3D spoiled gradient-echo sequence with fat suppression. The first three images in Fig.~\ref{Fig:RARE} are reconstructions from 800 k-space radial lines, corresponding to scans of about 2 minutes. \emph{Multi-Coil Nonuniform inverse Fast Fourier Transform (MCNUFFT)} refers to a simple inversion of the measurement operator without any regularization. \emph{DeCoLearn} refers to a CNN trained in a self-supervised fashion to remove artifacts from MCNUFFT images obtained from 1600 radial lines (scans of about 4 minutes)~\cite{Gan.etal2021}. While DeCoLearn offers excellent reconstruction performance when applied to 2000-line data (see \emph{Reference} in Fig.~\ref{Fig:RARE}), its performance degrades when applied \emph{without retraining} to 800-line data. \emph{Regularization by Artifact Removal (RARE)} is a variant of PnP that is obtained by simply replacing the AWGN denoiser by DeCoLearn~\cite{Liu.etal2020}. Note that RARE successfully adapts DeCoLearn to 800-line data without retraining, which is due to its ability to leverage DeCoLearn as an image prior.

Fig.~\ref{Fig:SIMBA} presents an application of PnP to bio-microscopy using the IDT instrument described in~\cite{Wu.etal2020}. RED-SD and SIMBA are both used to reconstruct a 3D algae sample of $1024 \times 1024 \times 25$ voxels from $b = 89$ high-resolution intensity measurements. Both algorithms use exactly the same forward model and the same DnCNN AWGN denoiser. The per-iteration memory complexity of RED-SD is about $75$ GBs, which includes the storage of the 3D complex-valued transfer functions for each illumination $\{\Abm_i\}$, all the measured intensity images $\{\ybm_i\}$, and the estimate of the desired image $\xbm$. By using mini-batches of size $p = 10$, SIMBA significantly reduces the per-iteration memory complexity to about $11$ GBs. Additionally, SIMBA has significant per-iteration computational advantage over RED-SD due to its usage of mini-batch gradients. Despite these memory and computational advantages of SIMBA, the results in Fig.~\ref{Fig:SIMBA} clearly show its comparable performance to RED-SD in terms of imaging quality.

\section{Future Directions}  \label{sec:future}

The idea of encapsulating prior information using algorithmic updates is a fertile area with much room for growth.  For PnP-ADMM and related methods there are questions about which denoisers provide guaranteed convergence, how to accelerate convergence, and how to manage the trade-offs between modularity and reconstruction quality.  These questions also apply to MACE, with additional questions about how to select hyperparameters for each agent, how to balance the contributions of multiple agents, and how to use agents that work in different spaces (e.g., sinogram domain and space domain).   And of course, there are many new applications to explore.  

\section{Conclusions} \label{sec:conclusions}

Since their introduction in 2013, PnP methods have become a standard tool for computational imaging. They've been used in a remarkably diverse range of applications in which they provide state-of-the-art performance. When they were introduced, they provided what was arguably the first practical approach to integrating learned models with imaging physics to solve inverse imaging problems. A significant factor in their rapid growth in popularity was the ease with which they can be implemented. Alternative approaches to achieving this goal have since emerged, and in some cases provide better reconstruction performance, but this is achieved at the expense of a potentially time-consuming and data-dependent application-specific training process. PnP and the multi-agent extension MACE are particularly powerful for contexts in which the forward model is not fixed or in which there is insufficient labeled problem-specific training data. 

\bibliographystyle{IEEEtran}

\begin{thebibliography}{10}
\providecommand{\url}[1]{#1}
\csname url@samestyle\endcsname
\providecommand{\newblock}{\relax}
\providecommand{\bibinfo}[2]{#2}
\providecommand{\BIBentrySTDinterwordspacing}{\spaceskip=0pt\relax}
\providecommand{\BIBentryALTinterwordstretchfactor}{4}
\providecommand{\BIBentryALTinterwordspacing}{\spaceskip=\fontdimen2\font plus
\BIBentryALTinterwordstretchfactor\fontdimen3\font minus
  \fontdimen4\font\relax}
\providecommand{\BIBforeignlanguage}[2]{{%
\expandafter\ifx\csname l@#1\endcsname\relax
\typeout{** WARNING: IEEEtran.bst: No hyphenation pattern has been}%
\typeout{** loaded for the language `#1'. Using the pattern for}%
\typeout{** the default language instead.}%
\else
\language=\csname l@#1\endcsname
\fi
#2}}
\providecommand{\BIBdecl}{\relax}
\BIBdecl

\bibitem{Parikh.Boyd2014}
N.~Parikh and S.~Boyd, ``Proximal algorithms,'' \emph{Foundations and Trends in
  Optimization}, vol.~1, no.~3, pp. 123--231, 2014.

\bibitem{boyd2011distributed}
S.~Boyd, N.~Parikh, E.~Chu, B.~Peleato, and J.~Eckstein, ``Distributed
  optimization and statistical learning via the alternating direction method of
  multipliers,'' \emph{Foundations and Trends in Machine Learning}, vol.~3,
  no.~1, 2011.

\bibitem{venkatakrishnan2013plug}
\BIBentryALTinterwordspacing
S.~V. Venkatakrishnan, C.~A. Bouman, and B.~Wohlberg, ``Plug-and-play priors
  for model based reconstruction,'' ECE Technical Reports, Purdue University,
  Tech. Rep. 448, 2013. [Online]. Available:
  \url{http://docs.lib.purdue.edu/ecetr/448}
\BIBentrySTDinterwordspacing

\bibitem{DabovBM3D07}
K.~Dabov, A.~Foi, V.~Katkovnik, and K.~Egiazarian, ``Image denoising by sparse
  {3-D} transform-domain collaborative filtering,'' \emph{IEEE Transactions on
  Image Processing}, vol.~16, no.~8, pp. 2080--2095, 2007.

\bibitem{Kamilov-2017}
U.~S. Kamilov, H.~Mansour, and B.~Wohlberg, ``A plug-and-play priors approach
  for solving nonlinear imaging inverse problems,'' \emph{IEEE Signal Process.
  Lett.}, vol.~24, no.~12, pp. 1872--1876, Dec 2017.

\bibitem{r1-beck2009fast}
A.~Beck and M.~Teboulle, ``A fast iterative shrinkage-thresholding algorithm
  for linear inverse problems,'' \emph{SIAM J. Imaging Sciences}, vol.~2,
  no.~1, pp. 183--202, 2009.

\bibitem{combettes2011proximal}
P.~L. Combettes and J.-C. Pesquet, ``Proximal splitting methods in signal
  processing,'' in \emph{Fixed-point algorithms for inverse problems in science
  and engineering}.\hskip 1em plus 0.5em minus 0.4em\relax Springer New York,
  2011, pp. 185--212.

\bibitem{FCI2022}
C.~A. Bouman, \emph{Foundations of Computational Imaging: A Model-Based
  Approach}.\hskip 1em plus 0.5em minus 0.4em\relax Philadelphia: Society for
  Industrial and Applied Mathematics, 2022.

\bibitem{schmidhuber2015deep}
J.~Schmidhuber, ``Deep learning in neural networks: An overview,'' \emph{Neural
  Networks}, vol.~61, pp. 85--117, 2015.

\bibitem{venkatakrishnan2013school}
S.~V. Venkatakrishnan, C.~A. Bouman, and B.~Wohlberg, ``Plug-and-play priors
  for model based reconstruction,'' in \emph{IEEE Global Conf.\ Signal
  Process.\ and Inf.\ Process. ({GlobalSIP})}, 2013, pp. 945--948.

\bibitem{Ono-2017}
S.~Ono, ``Primal-dual plug-and-play image restoration,'' \emph{IEEE
  Sig.~Proc.~Lett.}, vol.~24, pp. 1108--1112, 2017.

\bibitem{Gan.etal2021}
W.~Gan, Y.~Sun, C.~Eldeniz, J.~Liu, H.~An, and U.~S. Kamilov,
  ``Deformation-compensated learning for image reconstruction without ground
  truth,'' \emph{IEEE Trans. Med. Imag.}, 2022, doi:10.1109/TMI.2022.3163018.

\bibitem{Liu.etal2020}
J.~Liu, Y.~Sun, C.~Eldeniz, W.~Gan, H.~An, and U.~S. Kamilov, ``{RARE}: {I}mage
  reconstruction using deep priors learned without ground truth,'' \emph{IEEE
  J. Sel. Topics Signal Process.}, vol.~14, no.~6, pp. 1088--1099, 2020.

\bibitem{Teodoro.etal2019}
A.~M. Teodoro, J.~M. Bioucas-Dias, and M.~Figueiredo, ``A convergent image
  fusion algorithm using scene-adapted {G}aussian-mixture-based denoising,''
  \emph{IEEE Trans. Image Process.}, vol.~28, no.~1, pp. 451--463, Jan. 2019.

\bibitem{Sridhar2020}
V.~Sridhar, X.~Wang, G.~T. Buzzard, and C.~Bouman, ``Distributed iterative {CT}
  reconstruction using multi-agent consensus equilibrium,'' \emph{IEEE Trans.
  Comp. Imag.}, vol.~6, pp. 1153--1166, 2020.

\bibitem{Reehorst2019}
E.~T. {Reehorst} and P.~{Schniter}, ``Regularization by denoising:
  Clarifications and new interpretations,'' \emph{IEEE Trans. Comp. Imag.},
  vol.~5, no.~1, pp. 52--67, 2019.

\bibitem{Buzzard.etal2017}
G.~T. Buzzard, S.~H. Chan, S.~Sreehari, and C.~A. Bouman, ``Plug-and-play
  unplugged: {O}ptimization free reconstruction using consensus equilibrium,''
  \emph{SIAM J. Imaging Sci.}, vol.~11, no.~3, pp. 2001--2020, 2018.

\bibitem{scico-2022}
T.~Balke, F.~Davis, C.~Garcia-Cardona, M.~McCann, L.~Pfister, and B.~Wohlberg,
  ``{S}cientific {C}omputational {I}maging {CO}de {(SCICO)},'' Software library
  available from \url{https://github.com/lanl/scico}, 2022.

\bibitem{Meinhardt.etal2017}
T.~Meinhardt, M.~Moeller, C.~Hazirbas, and D.~Cremers, ``Learning proximal
  operators: {U}sing denoising networks for regularizing inverse imaging
  problems,'' in \emph{Proc. IEEE Int. Conf. Comp. Vis.}, Oct. 2017, pp.
  1799--1808.

\bibitem{Sun2019a}
Y.~Sun, B.~Wohlberg, and U.~S. Kamilov, ``An online plug-and-play algorithm for
  regularized image reconstruction,'' \emph{IEEE Trans. Comp. Imag.}, vol.~5,
  no.~3, pp. 395--408, Sep. 2019.

\bibitem{Bauschke2011}
H.~H. Bauschke and P.~L. Combettes, \emph{Convex Analysis and Monotone Operator
  Theory in Hilbert Spaces}, 2nd~ed.\hskip 1em plus 0.5em minus 0.4em\relax
  Springer, 2017.

\bibitem{sreehari2016TCI}
S.~Sreehari, S.~V. Venkatakrishnan, B.~Wohlberg, G.~T. Buzzard, L.~F. Drummy,
  J.~P. Simmons, and C.~A. Bouman, ``Plug-and-play priors for bright field
  electron tomography and sparse interpolation,'' \emph{IEEE Trans. Comp.
  Imag.}, vol.~2, no.~4, pp. 408--423, Dec 2016.

\bibitem{Chan2017}
S.~H. Chan, X.~Wang, and O.~A. Elgendy, ``Plug-and-play {ADMM} for image
  restoration: Fixed-point convergence and applications,'' \emph{IEEE Trans.
  Comp. Imag.}, vol.~3, no.~1, pp. 84--98, March 2017.

\bibitem{Ryu2019}
E.~Ryu, J.~Liu, S.~Wang, X.~Chen, Z.~Wang, and W.~Yin, ``Plug-and-play methods
  provably converge with properly trained denoisers,'' in \emph{Proc. 36th Int.
  Conf. Machine Learning (ICML)}, vol.~97, Jun. 2019, pp. 5546--5557.

\bibitem{Sun.etal2021}
Y.~Sun, Z.~Wu, X.~Xu, B.~Wohlberg, and U.~S. Kamilov, ``Scalable plug-and-play
  {ADMM} with convergence guarantees,'' \emph{IEEE Trans. Comp. Imag.}, vol.~7,
  pp. 849--863, Jul. 2021.

\bibitem{Sun.etal2019b}
Y.~Sun, J.~Liu, and U.~S. Kamilov, ``Block coordinate regularization by
  denoising,'' in \emph{Proc. Advances in Neural Information Processing Systems
  33}, Vancouver, BC, Canada, Dec. 2019, pp. 382--392.

\bibitem{Xu.etal2020}
X.~Xu, Y.~Sun, J.~Liu, and U.~S. Kamilov, ``Provable convergence of
  plug-and-play priors with {MMSE} denoisers,'' \emph{IEEE Signal Process.
  Lett.}, vol.~27, pp. 1280--1284, 2020.

\bibitem{Liu.etal2021b}
J.~Liu, S.~Asif, B.~Wohlberg, and U.~S. Kamilov, ``Recovery analysis for
  plug-and-play priors using the restricted eigenvalue condition,'' in
  \emph{Proc. Advances in Neural Information Processing Systems 35}, December
  6-14, 2021.

\bibitem{Romano2017}
Y.~Romano, M.~Elad, and P.~Milanfar, ``The little engine that could:
  Regularization by denoising {(RED)},'' \emph{SIAM J. Imaging Sciences},
  vol.~10, no.~4, pp. 1804--1844, 2017.

\bibitem{Zhang.Ghanem2018}
J.~Zhang and B.~Ghanem, ``{ISTA}-{Net}: {I}nterpretable optimization inspired
  deep network for image compressive sensing,'' in \emph{Proc. {IEEE} Conf.
  Comp. Vision and Pattern Recog. ({CVPR})}, 2018, pp. 1828--1837.

\bibitem{Wu.etal2020}
Z.~Wu, Y.~Sun, A.~Matlock, J.~Liu, L.~Tian, and U.~S. Kamilov, ``{SIMBA}:
  {S}calable inversion in optical tomography using deep denoising priors,''
  \emph{IEEE J. Sel. Topics Signal Process.}, vol.~14, no.~6, pp. 1163--1175,
  2020.

\bibitem{r2-gregor2010learning}
K.~Gregor and Y.~LeCun, ``Learning fast approximation of sparse coding,'' in
  \emph{Proc. 27th Int. Conf. Machine Learning (ICML)}, Haifa, Israel, June
  21-24, 2010, pp. 399--406.

\bibitem{r3-chen_theoretical_2018}
X.~Chen, J.~Liu, Z.~Wang, and W.~Yin, ``Theoretical linear convergence of
  unfolded {ISTA} and its practical weights and thresholds,'' in \emph{Proc.
  Advances in Neural Information Processing Systems 31}, 2018, pp. 9079--9089.

\bibitem{Monga.etal2021}
V.~Monga, Y.~Li, and Y.~C. Eldar, ``Algorithm unrolling: {I}nterpretable,
  efficient deep learning for signal and image processing,'' \emph{IEEE Signal
  Process. Mag.}, vol.~38, no.~2, pp. 18--44, Mar. 2021.

\bibitem{Gilton.etal2021}
D.~Gilton, G.~Ongie, and R.~Willett, ``Deep equilibrium architectures for
  inverse problems in imaging,'' \emph{IEEE Trans. Comp. Imag.}, vol.~7, pp.
  1123--1133, Oct. 2021.

\bibitem{r4-zoran2011learning}
D.~Zoran and Y.~Weiss, ``From learning models of natural image patches to whole
  image restoration,'' in \emph{Proc. IEEE Int. Conf. Comp. Vis. (ICCV)},
  Barcelona, Spain, Nov. 2011, pp. 479--486.

\bibitem{r5-zhang2017learning}
K.~Zhang, W.~Zuo, S.~Gu, and L.~Zhang, ``Learning deep {CNN} denoiser prior for
  image restoration,'' in \emph{Proc. {IEEE} Conf. Computer Vision and Pattern
  Recognition ({CVPR})}, Jul. 2017, pp. 2808--2817.

\bibitem{r6-tirer2019image}
T.~Tirer and R.~Giryes, ``Image restoration by iterative denoising and backward
  projections,'' \emph{IEEE Trans. Image Process.}, vol.~28, no.~3, pp.
  1220--1234, 2019.

\bibitem{r7-tirer2020back}
------, ``Back-projection based fidelity term for ill-posed linear inverse
  problems,'' \emph{IEEE Trans. Image Process.}, vol.~29, pp. 6164--6179, 2020.

\bibitem{r8-arjomand_bigdeli_deep_2017}
S.~A. Bigdeli, M.~Jin, P.~Favaro, and M.~Zwicker, ``Deep mean-shift priors for
  image restoration,'' in \emph{Proc. Advances in Neural Information Processing
  Systems 30}, Dec 4-9, 2017, pp. 763--772.

\bibitem{r9-alain2014what}
G.~Alain and Y.~Bengio, ``What regularized auto-encoders learn from the
  data-generating distribution,'' \emph{J. Mach. Learn. Res.}, vol.~15, no.~1,
  p. 3563–3593, jan 2014.

\bibitem{r10-teodoro2016image}
A.~M. Teodoro, J.~M. Biocas-Dias, and M.~A.~T. Figueiredo, ``Image restoration
  and reconstruction using variable splitting and class-adapted image priors,''
  in \emph{Proc. {IEEE} Int. Conf. Image Proc. ({ICIP} 2016)}, September 25-28,
  2016, pp. 3518--3522.

\bibitem{r11-teodoro2019convergent}
A.~M. Teodoro, J.~M. Bioucas-Dias, and M.~A.~T. Figueiredo, ``A convergent
  image fusion algorithm using scene-adapted gaussian-mixture-based
  denoising,'' \emph{IEEE Trans. Image Process.}, vol.~28, no.~1, pp. 451--463,
  2019.

\bibitem{r12-tirer2019super}
T.~Tirer and R.~Giryes, ``Super-resolution via image-adapted denoising cnns:
  Incorporating external and internal learning,'' \emph{IEEE Signal Process.
  Lett.}, vol.~26, no.~7, pp. 1080--1084, 2019.

\bibitem{gnanasambandam-2020-one}
A.~Gnanasambandam and S.~Chan, ``One size fits all: Can we train one denoiser
  for all noise levels?'' in \emph{Proc. 37th Int. Conf. Machine Learning
  (ICML)}, vol. 119, 13--18 Jul 2020, pp. 3576--3586.

\bibitem{r13-Shocher_2018_CVPR}
A.~Shocher, N.~Cohen, and M.~Irani, ````{Zero-shot}'' super-resolution using
  deep internal learning,'' in \emph{Proc. {IEEE} Conf. Comp. Vision and
  Pattern Recog. ({CVPR})}, June 2018.

\bibitem{r14-9156619}
S.~A. Hussein, T.~Tirer, and R.~Giryes, ``Correction filter for single image
  super-resolution: Robustifying off-the-shelf deep super-resolvers,'' in
  \emph{Proc. {IEEE} Conf. Comp. Vision and Pattern Recog. ({CVPR})}, 2020, pp.
  1425--1434.

\bibitem{majee2021}
S.~Majee, T.~Balke, C.~A.~J. Kemp, G.~T. Buzzard, and C.~A. Bouman,
  ``Multi-slice fusion for sparse-view and limited-angle {4D CT}
  reconstruction,'' \emph{IEEE Trans. Comp. Imag.}, vol.~7, pp. 448--462, 2021.

\bibitem{Reid2022}
E.~J. Reid, L.~F. Drummy, C.~A. Bouman, and G.~T. Buzzard, ``Multi-resolution
  data fusion for super resolution imaging,'' \emph{IEEE Trans. Comp. Imag.},
  vol.~8, pp. 81--95, 2022.

\bibitem{Pellizzari2020}
C.~J. Pellizzari, M.~F. Spencer, and C.~A. Bouman, ``Coherent plug-and-play:
  Digital holographic imaging through atmospheric turbulence using model-based
  iterative reconstruction and convolutional neural networks,'' \emph{IEEE
  Trans. Comp. Imag.}, vol.~6, pp. 1607--1621, 2020.

\end{thebibliography}

\footnotesize

\bigskip
\noindent
\textbf{Ulugbek S. Kamilov} (SM, kamilov@wustl.edu) received the B.Sc./M.Sc.\ degree in communication systems and the Ph.D.\ degree in EE from EPFL, Lausanne, Switzerland, in 2011 and 2015, respectively. He is currently an Assistant Professor and the Director of Computational Imaging Group with Washington University in St.\ Louis. From 2015 to 2017,
he was a Research Scientist with Mitsubishi Electric Research Laboratories. He is the recipient of an NSF CAREER Award and the IEEE Signal Processing Society’s 2017 Best Paper Award. 

\medskip
\noindent
\textbf{Charles Bouman}
(F, bouman@purdue.edu) received a B.S.E.E. degree from the University of Pennsylvania in 1981 
and a MS degree from the UC Berkeley in 1982. 
From 1982 to 1985, he was a full staff member at MIT Lincoln Laboratory 
and in 1989 he received a Ph.D. in EE from Princeton University. 
He joined Purdue University in 1989 where he is currently the Showalter Professor 
of Electrical and Computer Engineering and Biomedical Engineering. 
Professor Bouman is a Fellow of the IEEE, an Honorary Member of the IS\&T and a Fellow of the National Academy of Inventors (NAI).
He is a recipient of the IEEE Signal Processing Society, 2021 Claude Shannon-Harry Nyquist Technical Achievement Award and a co-recipient of the first place award for the 2022 AAPM TrueCT Reconstruction Challenge.

\medskip
\noindent
\textbf{Gregery Buzzard}
 (SM, buzzard@purdue.edu) 
received degrees in violin performance, computer science, and mathematics from Michigan State University, and received the Ph.D. degree in mathematics from the University of Michigan in 1995.  He held postdoctoral positions at Indiana University and Cornell University before joining the mathematics faculty at Purdue University in 2002, where he served as Department Head from 2013-2020.  He is a co-recipient of the first place award for the 2022 AAPM TrueCT Reconstruction Challenge.

\medskip
\noindent
\textbf{Brendt Wohlberg}
(SM, brendt@ieee.org) received the BSc (Hons) degree in applied mathematics, and the MSc and PhD degrees in electrical engineering from the University of Cape Town, South Africa, in 1990, 1993, and 1996, respectively. He is currently a Staff Scientist in Theoretical Division at Los Alamos National Laboratory. He was a co-recipient of the 2020 SIAM Activity Group on Imaging Science Best Paper Prize, was previously Editor-in-Chief of the IEEE Transactions on Computational Imaging, and is currently Editor-in-Chief of the IEEE Open Journal of Signal Processing.

\end{document}